\documentclass[conference]{IEEEtran}
\IEEEoverridecommandlockouts

\usepackage{cite}
\usepackage{amsmath,amssymb,amsfonts}
\usepackage{algorithmic}
\usepackage{xcolor}
\usepackage{graphicx}
\usepackage{hyperref}
\usepackage{textcomp}
\usepackage{multirow}
\usepackage{url}
\usepackage{amsmath}
\def\BibTeX{{\rm B\kern-.05em{\sc i\kern-.025em b}\kern-.08em
    T\kern-.1667em\lower.7ex\hbox{E}\kern-.125emX}}
\begin{document}

\title{Impression Zombies: Characteristics Analysis and Classification of New Harmful Accounts on Social Media\\
}

\author{\IEEEauthorblockN{Keito Uehara}
\IEEEauthorblockA{\textit{Yokohama National University} \\
Kanagawa, Japan \\
uehara-keito-jz@ynu.jp}
\and
\IEEEauthorblockN{Taichi Murayama}
\IEEEauthorblockA{\textit{Yokohama National University} \\
Kanagawa, Japan \\
murayama-taichi-bs@ynu.ac.jp}

}

\maketitle

\begin{abstract}
``Impression Zombies'', a type of malicious account designed to artificially inflate engagement metrics, have recently emerged as a significant threat on X (formerly Twitter). These accounts disseminate a high volume of low-quality, irrelevant posts, which degrade the user experience. This study aims (1) to quantitatively characterize their behavioral patterns and (2) to develop a method for detecting such accounts. To address the first objective, we collected data from 9,909 accounts and compared the characteristics of Impression Zombies and general users within this dataset. We find that, Impression Zombies post more than three times the average total number of posts per day and tend to gather followers by using phrases such as ``follow back.'' Addressing the second objective, we constructed a classification model for Impression Zombies that leverages the contextual incoherence often observed between parent posts and the replies from Impression Zombies. Experimental results show that our model achieved approximately 92\% accuracy in detecting Impression Zombies. This study provides the first quantitative insights into Impression Zombies and offers a practical framework for detecting such accounts, contributing to platform transparency and the health of social media ecosystems.

\end{abstract}


\section{Introduction}
\label{sec_intro}
The proliferation of social media has established platforms like X (formerly Twitter) as major infrastructure for real-time news and public discourse. However, concerns about the integrity of its information circulation are increasing. As one example, a group of accounts referred to as ``Impression Zombies'' has become a significant issue within the Japanese-speaking community, since X launched a program in July 2023 to distribute a portion of its advertising revenue to creators, ~\cite{intro1}. Under this revenue program, the amount of money creators can receive increases with the number of views or impressions their posts garner. To maximize these impressions and their revenue, Impression Zombies engage in disruptive tactics, such as posting nonsensical replies to high-profile posts, creating posts that merely list the top trending words, and duplicating the posts of others. Such behavior hinders general users from accessing the necessary information. For instance, during the magnitude 7.6 earthquake that struck Ishikawa Prefecture in Japan on January 1, 2024, Impression Zombies exploited the disaster to farm engagement, spreading fake rescue requests and unrelated images that misled users~\cite{intro2}. Although Impression Zombies have become a cause of social problems, their actual behavioral mechanisms and characteristics have not yet been sufficiently clarified.

Related phenomena on social media include Clickbait~\cite{jung2022click,lischka2023clickbait} and Engagement Bait~\cite{watti2023avoid}. These are similar to Impression Zombies in that they attract attention by frequently posting. However, a key distinction lies in their underlying objectives. While Engagement Bait and Clickbait primarily aim to drive user traffic to external websites or advertisements.In contrast, Impression Zombies aim to maximize the visibility of their own posts. This difference in purpose is reflected in the surface-level characteristic that Impression Zombies posts rarely contain external links, whereas Engagement Bait and Clickbait consistently attempt to guide users off-platform. Furthermore, Impression Zombies often use other languages, such as English or Hindi, despite operating within the Japanese-speaking community. These unique linguistic and behavioral patterns suggest that Impression Zombies constitute a novel class of malicious accounts, whose behaviors differ significantly from conventional harmful accounts. However, their specific behavioral patterns and characteristics remain largely unexamined, presenting a clear research gap.

\begin{figure}[t]
    \centering
    \includegraphics[width=1.0\linewidth]{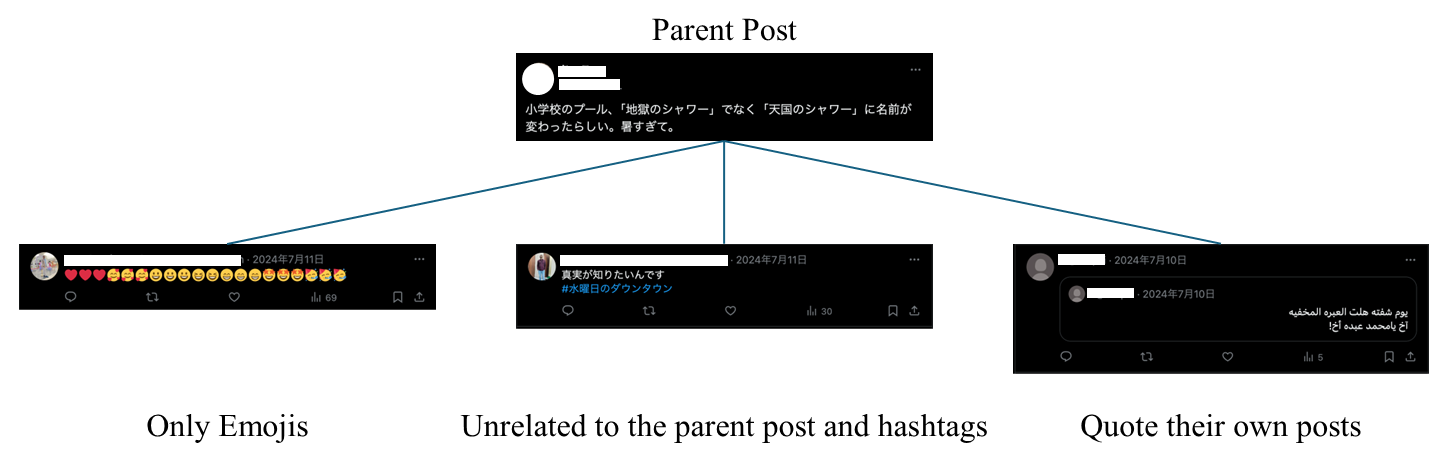}
    \caption{Examples of replies lacking contextual connection to the parent post and the corresponding Impression Zombies.}
    \label{fig:example}
\end{figure}

To address this gap, this study makes two primary contributions. First, we provide the first quantitative characterization of Impression Zombies. We hypothesize that their activity patterns differ significantly from general users, and verify this through a comparative analysis of account-level metadata such as account age and posting frequency. Second, we develop a classification model to distinguish Impression Zombies from general users. The model utilizes a characteristic property of their replies, which is the lack of a semantic connection to the parent post~\ref{fig:example}. By leveraging this contextual incongruity, the model achieved a classification accuracy of approximately 92\%. This paper aims to enhance the understanding of monetization-driven user behaviors and improve the integrity of social media platforms by quantitatively clarifying the behavioral characteristics of Impression Zombies and proposing a novel classification method utilizing both post and reply texts.

\section{Related Work}
``Impression Zombies,'' also known by terms like ``Engagement Farming'' or ``Impression Farmer,'' is not unique to Japan and has emerged as a pervasive issue in global social media communities. Here, this section illustrate that Impression Zombies represent a new category of accounts with distinct features. By examining examples of various accounts that contaminate the social media information ecosystem, we show that Impression Zombies possess characteristics that differentiate them from these pre-existing types.

\subsection{Spam Accounts}
Spam accounts are a primary example of accounts that adversely affect social media. These accounts engage in malicious activities such as promoting products and services~\cite{bazzaz2021hybrid}, artificially inflating follower counts~\cite{ferrara2022twitter}, indiscriminately sending unsolicited messages to users~\cite{alsaleh2015combating}, and directing them to phishing sites~\cite{bazzaz2021hybrid}. Within this category, bots are a specific subset controlled by automated programs designed to mimic human behavior~\cite{feng2021twibot,lei2022bic}. Bots have been utilized to spread disinformation across a wide range of subjects~\cite{cresci2018fake,nizzoli2020charting,yang2013empirical}, including public health topics such as the COVID-19 pandemic~\cite{shi2020social,ferrara2020types} and the manipulation of public opinion during major political events~\cite{bessi2016social, bastos2019brexit,gorodnichenko2021social,Ferrara_2017}.

While spam and bots operate with explicit malicious intent, Impression Zombies have a different primary objective: to maximize the visibility of their posts. Impression Zombies typically replicate trending posts or post nonsensical replies, distinguishing themselves as a group of accounts with characteristics different from spam and bots that disseminate information with clear malicious intent. Therefore, detecting Impression Zombies requires the construction of a new, specialized dataset adapted for them, rather than relying on existing datasets for spam and bots detections.

\subsection{AI-generated harmful content}
With the proliferation and advancement of AI technology, ``AI-Slop''~\cite{klincewicz2025slopaganda} and deepfake~\cite{pawelec2022deepfakes,drolsbach2025characterizing}, are being mass-produced on an unprecedented scale. This surge in synthetic media has made it increasingly difficult for users to distinguish truth from falsehood. Similar to Impression Zombies, this existence poses a new threat to the integrity of information circulation on social media. However, accounts that create and disseminate AI-Slop and deepfakes are distinct from Impression Zombies in that the former serve as the origin of the content, creating content from scratch, whereas the latter merely swarm or replicate the posts of others.

\subsection{Clickbait and Engagement Bait}
Similar to Impression Zombies, Clickbait~\cite{jung2022click,lischka2023clickbait} and Engagement Bait~\cite{watti2023avoid} are types of existence that also aim to maximize the visibility of their articles. They are characterized by the use of sensational or eye-catching headlines for posts and articles to capture the attention of a wide audience. Their objective is to entice users with these headlines to click on links leading to external websites or advertisements. In contrast, the purpose of Impression Zombies are strictly limited to having their posts seen within the X platform itself. As their actions are confined to gaining engagement on X, Impression Zombies constitute a distinct group of accounts from Clickbait and Engagement Bait, which are designed to drive traffic to external sites.

\section{Datasets}
\label{sec_datasets}

We constructed these dataset derived from a primary Impression Zombies dataset: an \textit{Account dataset} and \textit{Classifcation dataset}.  \textit{Accounts Dataset} is utilized for the behavioral features analysis of Impression Zombies, while \textit{Classification Dataset} is used to train and evaluate our classification model. Additionally, a \textit{Parent-Reply Pair Dataset} is utilized fine-tuning for classification model.

\subsection{Impression Zombies Dataset}
During the period from July 11 to 18, 2024, we used the X v2 API to collect a total of 9,909 replies to 101 Japanese posts that had each gained more than 50,000 likes within 24 hours of posting. From this data, we constructed two datasets: Account Dataset and Classification Dataset.

\subsubsection{Account dataset}
Account dataset contains user information from X, with each user assigned either a general user or Impression Zombies label, to conduct comparison analysis. In this paper, the "general user" category represents any user not classified as an Impression Zombie. We attempted to retrieve user profile information for all 6,278 accounts. Of these, 5,497 accounts were successfully obtained, while the remaining 781 were already suspended or deleted at the time of collection. 

To assign a general user or Impression Zombies label to a user, we established the following four judgment criteria.
\begin{itemize}
    \item Screen Name: An account with screen names that appeared non-Japanese was considered a potential Impression Zombies.

    \item Verification Mark: An account bearing a verification mark was considered a potential Impression Zombies. This mark signifies a subscription to X's premium services, which is a prerequisite for receiving revenue.

    \item profile sentence: An account with a profile written in a language other than Japanese was considered a potential Impression Zombies.

    \item Reply Content: An account with their replies that disconnected from the context of the parent post and employed frequent use of formulaic or bot-like expressions was considered a potential Impression Zombies.
    
\end{itemize} 
After the annotation process, the dataset consisted of 4,504 general users and 993 Impression Zombies.

\subsubsection{Classification dataset}
Classification dataset consists of 9,909 pairs of Japanese parent posts and replies. Each reply was labeled as either an Impression Zombies or a general user by four annotators, including the authors. A reply was labeled as originating from an Impression Zombies if two or more of the four annotators judged it on the following criteria:  (1) the reply was semantically nonsensical or irrelevant to the parent post, and (2) the language of the reply was mechanical, as if generated by AI or a machine translation. As a result, a dataset was constructed containing 5,549 general user replies and 4,360 Impression Zombies replies. Inter-annotator agreement, Fleiss' kappa score, was 0.79, confirming the reliability of the evaluations.

\subsection{Parent-Reply Pair Dataset}
For fine-tuning our classification model, we prepared a clean dataset of one-to-one 250,000 pairs of Japanese parent posts and their replies collected between January and March 2020, prior to the emergence of Impression Zombies (July 2023). This corpus contains standard user interactions and does not include Impression Zombies.

\section{Characteristics Analysis of Impression Zombies}
\label{sec_analysis}
Impression Zombies are emerging type of user on X whose behavioral mechanisms and structural attributes remain largely unexplored. Therefore, using the \textit{Accounts dataset}, we conduct a quantitative analysis of the characteristics observed in Impression Zombies in comparison to general users. We quantitatively analyze four key aspects:
(1) the relationship between the number of days since account creation (account age) and the total number of posts,
(2) linguistic characteristics of profile sentences,
(3) the ratio of following to follower count,
and (4) temporal posting patterns.

\subsection{Users' Account Age and Total Posts}

\begin{figure*}[t]
    \centering
    \includegraphics[width=1.0\linewidth]{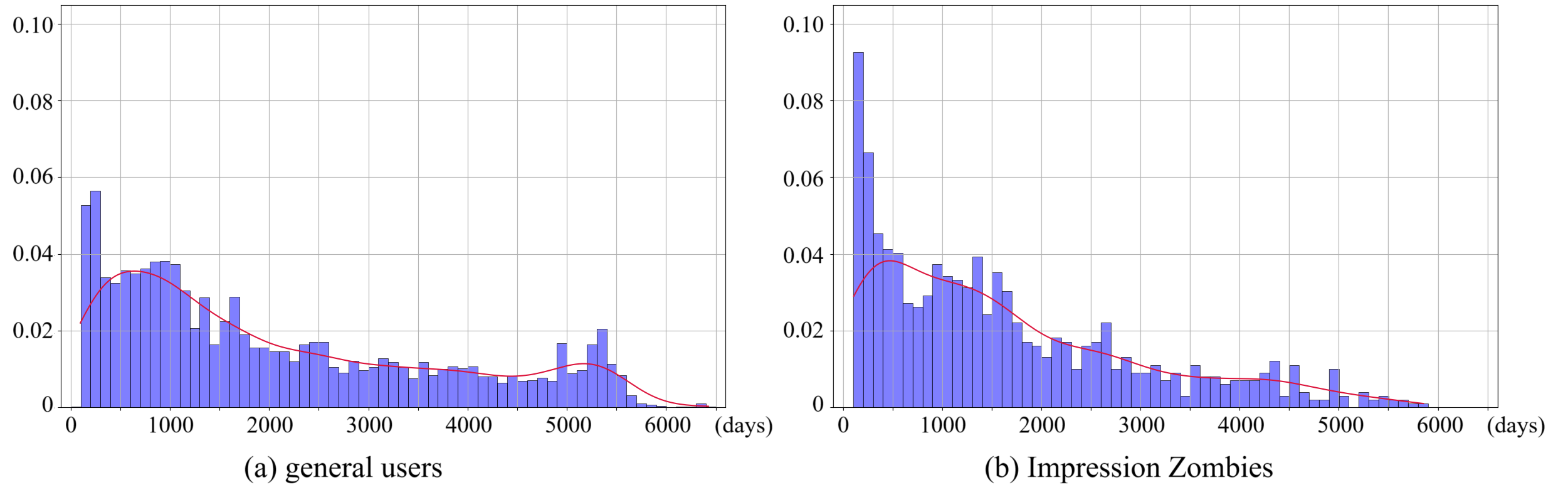}
    \caption{Histogram of account age measured from the account creation date to November 1, 2024. (a) shows the distribution of general users, and (b) shows the distribution of Impression Zombies. The x-axis represents elapsed days, and the y-axis represents the relative frequency. The red line indicates the Kernel Density Estimate (KDE) curve. The difference in distribution between general users and Impression Zombies were statistically significant, with a t-test yielding a p-value below 0.001.}
    \label{fig:age}
\end{figure*}

\begin{figure*}[t]
    \centering
    \includegraphics[width=1.0\linewidth]{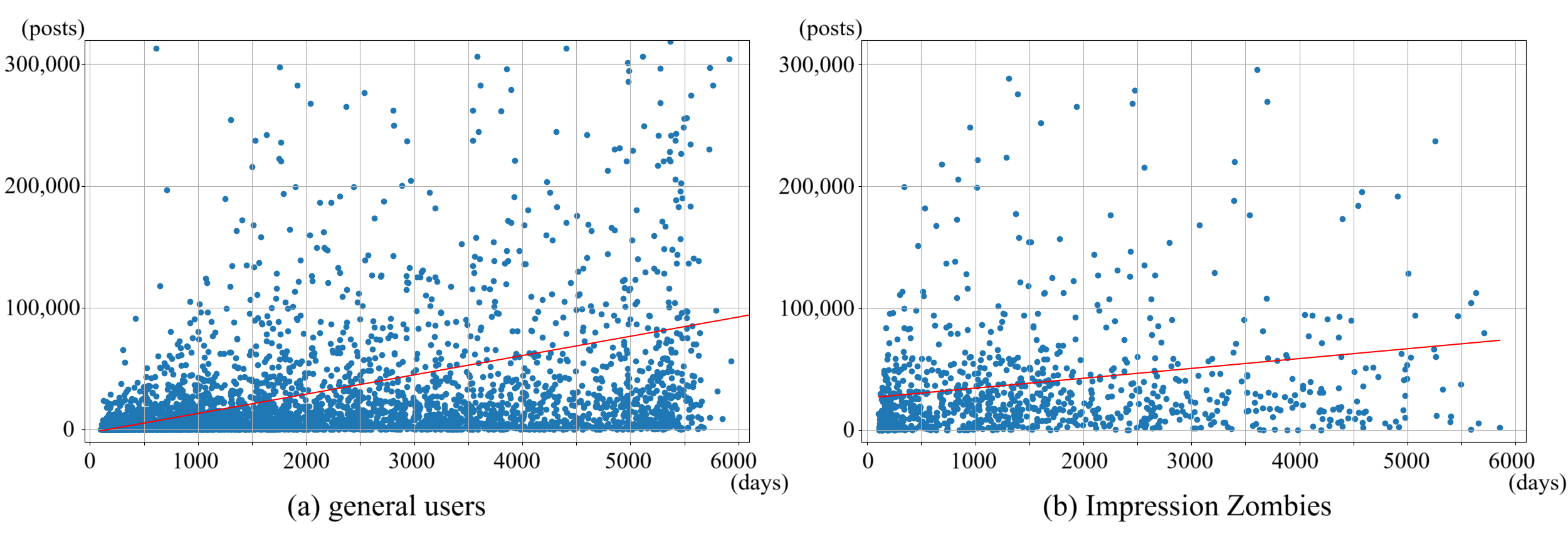}
    \caption{Scatter plot of the total number of posts versus account age (days from creation to November 1, 2024) for (a) general users and (b) Impression Zombies. The x-axis represents the elapsed days, and the y-axis represents the total number of posts. The red line indicates the regression line. The difference in distribution between general users and Impression Zombies were statistically significant, with a t-test yielding a p-value below 0.001.}
    \label{fig:age_vs_posts}
\end{figure*}

As depicted in Figure~\ref{fig:age}, the distribution of account ages shows a notable difference between Impression Zombies and general users. Specifically, for accounts older than 1,800 days (approx. 5 years), general users constitute 44\% of the cohort, whereas Impression Zombies constitute only 34\%. On the other hand, for newer accounts (under 500 days old), general users constitute about 18\%, whereas Impression Zombies constitute approximately 25\%. This 500-day period roughly corresponds to the time from July 2023, when X launched its ad revenue sharing program for creators. This suggests that Impression Zombies tend to be newer than those of general users, likely because their operators created them with the specific intention of generating revenue.

Figure~\ref{fig:age_vs_posts} shows the relationship between account age and total posts. It reveals that users with over 10,000 total posts constitute 41\% of general users, compared to about 72\% of Impression Zombies. This disparity becomes even more pronounced when examining the average daily posting rate. General users averaged 13.98 posts per day, whereas Impression Zombies averaged 42.88 posts, which is over three times higher.
In particular, when focusing specifically on newer accounts (under 500 days old), only about 11\% of general users had more than 10,000 posts, compared to approximately 60\% of Impression Zombies. The average number of daily posts for this group also showed a stark difference: 13.58 for general users versus 87.16 for Impression Zombies. This pattern suggests that Impression Zombies are posting an exceptionally high volume of content in a short period, exceeding that of general users.

\subsection{Profile Sentence}
\label{subsec:description}

\begin{table}[t]
    \centering
    \caption{Top 10 of characteristic bi-grams ordered by their odds ratios in the profile descriptions of Impression Zombies. Odds ratios were calculated using logistic regression.}
    
    \begin{tabular}{c|c}
       bi-gram  &  odds ratio\\
    \hline
       follow back  & 18.09\\
       dm for & 11.37\\
       social activist & 10.42\\
       jay shree & 9.93\\
       content creator & 8.98\\
       no dm & 8.56\\
       govt teacher & 8.27\\
       believe in & 7.66\\
       cricket lover & 7.44\\
       social worker & 5.70\\
    \hline

    \end{tabular}
    \label{tab:description}
\end{table}

To identify the linguistic patterns in profile descriptions that distinguish Impression Zombies from general users, we conducted a logistic regression analysis. The dependent variable was a binary label assigned to each profile: 1 for an Impression Zombies and 0 for a general user. The analysis yielded an odds ratio for each bi-gram, where a value greater than one indicates that the bi-gram's presence increases the likelihood of an account being an Impression Zombies. Table~\ref{tab:description} lists the top 10 bi-grams most characteristic of Impression Zombies, ordered by their odds ratios, are listed in Table~\ref{tab:description}.

The prevalence of the bi-gram ``follow back'' strongly indicates that Impression Zombies are attempting to increase their follower count, likely to meet the eligibility criteria for X's monetization program or to expand their audience. Indeed, some Impression Zombies had several thousand to tens of thousands of followers. As there is little reason for general users to follow Impression Zombies, we infer that they use the phrase ``follow back'' to increase their follower count.
The second-ranked bi-gram , often co-occurred with terms related to financial solicitation, such as ``ads,'' ``promotions,'' and ``business'', suggesting that the activities of Impression Zombies are driven by financial incentives. Furthermore, the terms ``social activist,'' ``content creator,'' and ``social worker'' are likely included as part of a self-branding strategy to appeal to other users by presenting themselves as actively engaged accounts. These findings that Impression Zombies strategically craft their profiles to achieve their primary objective maximizing the visibility of their posts to a broad audience.

\subsection{Following and Followers}
\label{subsec:ffratio}

\begin{figure*}[t]
    \centering
    \includegraphics[width=1.0\linewidth]{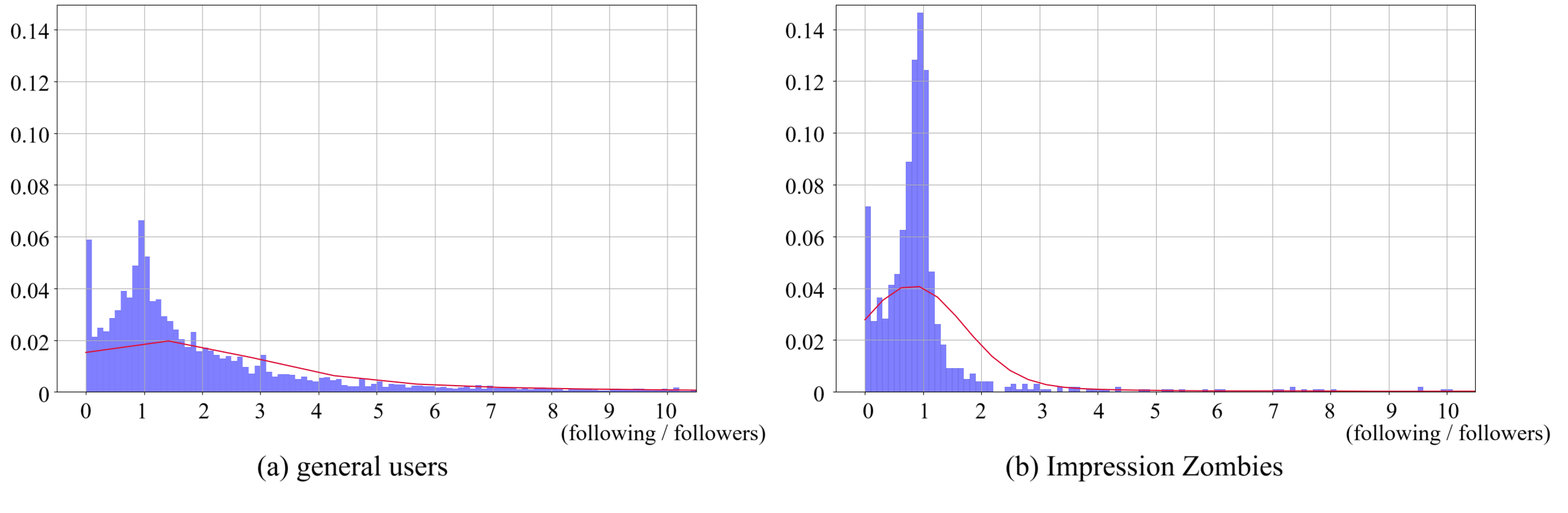}
    \caption{Distribution of the following-to-follower ratio for (a) general users and (b) Impression Zombies. The x-axis represents this ratio, and the y-axis represents the relative frequency. The red line indicates the KDE curve. The average following-to-follower ratio was 2.93 for general users, in contrast to 1.26 for Impression Zombies. The difference in distribution between general users and Impression Zombies were statistically significant, with a t-test yielding a p-value below 0.001.}
    \label{fig:ffratio}
\end{figure*}

Figure~\ref{fig:ffratio} shows the distribution of the following-to-follower ratio (the number of accounts followed divided by the number of followers), where a value greater than one indicates that an account follows more users than it has followers. Among users with ratios within 0.10 of 1.00, general users accounted for approximately 15\%, whereas Impression Zombies accounted for about 32\%. On the other hand, users with a ratio of 2.0 or higher constituted about 35\% of general users, but only about 7\% of Impression Zombies. In other words, while general users tend to follow significantly more accounts than they have followers, Impression Zombies tend to have a following count that is close to their follower count. Such balance is likely attributable to the fact that the profile sentences of Impression Zombies often contain the phrase ``follow back'' (discussed in Section~\ref{subsec:description}). Furthermore, This reciprocal pattern resembles following loops previously observed among social bots. Moreover, it is possible that other mechanisms, such as follower-buying schemes, are also contributing factors.

\subsection{Timestamp}
\label{subsec:heatmap}

\begin{figure*}[t]
    \centering
    \includegraphics[width=1.0\linewidth]{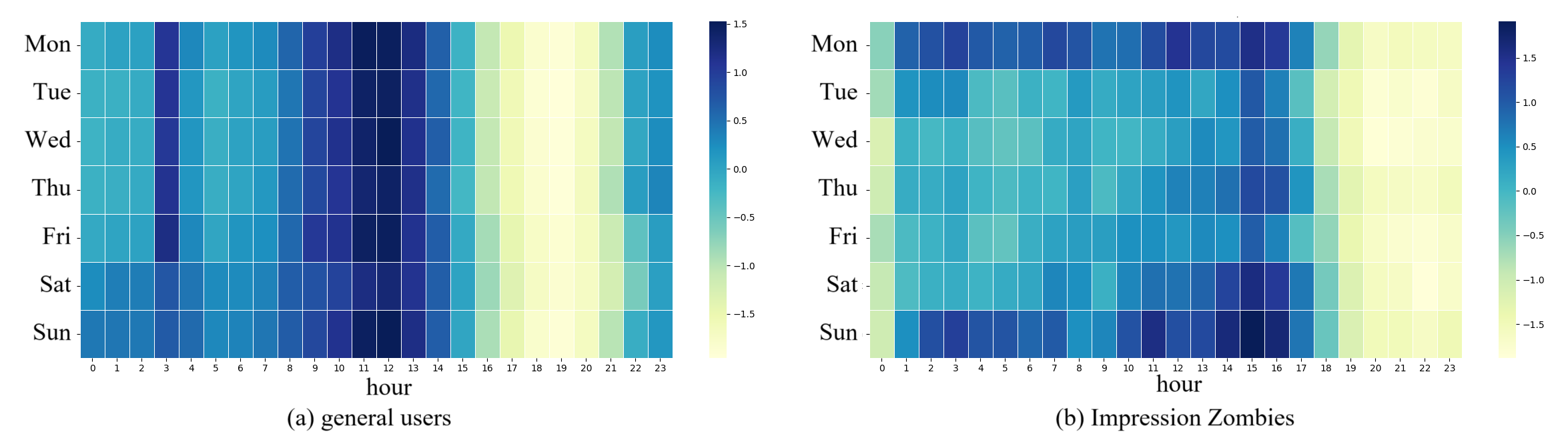}
    \caption{Heatmap showing the number of posts made at different times of the day for (a) general users and (b) Impression Zombies. The x-axis shows the time of day and the y-axis shows the day of the week. Color intensity represents high (purple) or low (yellow) activity.}
    \label{fig:heatmap}
\end{figure*}

Figure~\ref{fig:heatmap} shows a heatmap illustrating the number of posts by day of the week and time of day for (a) general users and (b) Impression Zombies. Each cell value represents the standardized Z-score of the hourly post counts. A purple cell indicates that the number of posts in a given time slot is significantly high, while a yellow is significantly low. Heatmap shows that the peak posting time for (b) Impression Zombies are 15:00, lagging approximately three hours behind the 12:00 peak for general users. In terms of weekly activity, their post volume tends to be concentrated on Mondays and Sundays. The consistency of these daily and weekly patterns suggests that the activity is not fully automated but likely involves strategic human intervention.

\section{Classification for Impression Zombies' Post}
\label{sec_classification}
This section addresses the task of classification Impression Zombies from general users. We construct and evaluate a classification model that utilizes pairs of a parent post and its corresponding reply.

\begin{figure*}[t]
    \centering
    \includegraphics[width=1.0\linewidth]{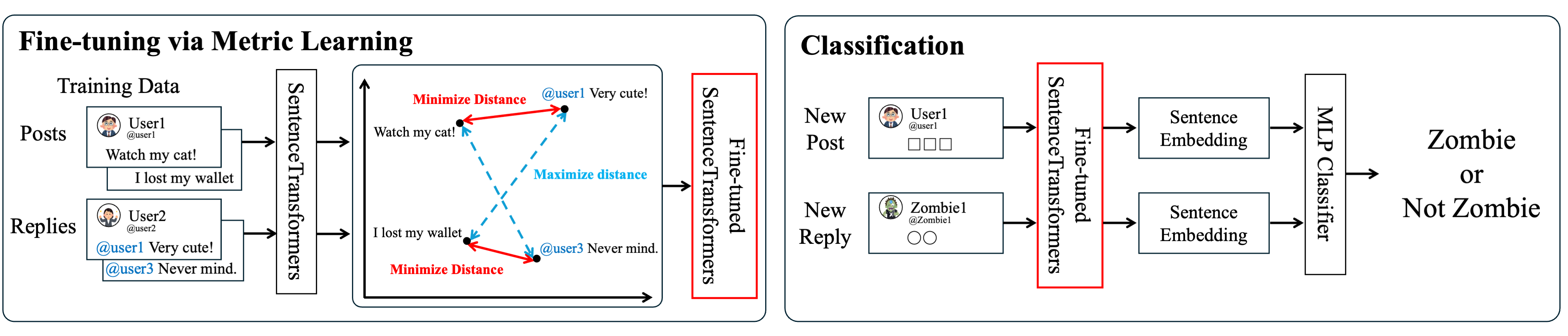}
    \caption{Overview of the proposed classification model, consisting of a fine-tuning phase and a classification phase. First, in the fine-tuning phase, we employed a metric learning approach to fine-tune a SentenceTransformers model. This process trains the model to construct an embedding space that captures the contextual relationship between posts and replies by pulling the vector representations of corresponding pairs closer together and pushing those of non-corresponding pairs further apart. Second, in the classification phase, we used the fine-tuned SentenceTransformers to encode each post-reply pair into a single vector. This vector was then entered into a multi-layer perceptron (MLP) classifier to determine whether the author of the reply was Impression Zombies or general users.}
    \label{fig:model}
\end{figure*}

\subsection{Proposed Classification Model}
\subsubsection{Hypothesis and Overall Framework}
Posts from Impression Zombies are often difficult to identify based on their text alone, as they frequently contain copied content from other posts or formulaic phrases. To address this limitation, we hypothesized that replies from Impression Zombies, which often consist of irrelevant character strings or only emojis, do not maintain a contextual relationship with the parent post. This contextual incoherence can therefore serve as a powerful clue for their classification. 

Based on this hypothesis, we construct a classification model that explicitly captures the semantic relationship between a parent post and its reply. We employ SentenceTransformers~\cite{reimers-gurevych-2019-sentence} to encode the text of the parent post and its corresponding reply into embedding representations. The concatenated embedding is then input into a MLP classifier to distinguish between replies from general users and Impression Zombies. Furthermore, to enhance the classification accuracy, we propose a method for Impression Zombies by fine-tuning the language model to consider the similarity between a parent post and its reply. An overview of the framework is shown in Figure~\ref{fig:model}.

\subsubsection{Fine-tuning of Embedding Model for Contextual Relationship}
\label{subsubsec:fine-tuning}
Our method is based on the hypothesis that Impression Zombie replies have significantly lower semantic similarity to their parent posts compared to genuine replies. We posit that fine-tuning the embedding space to explicitly reflect this distinction will enhance classification accuracy. To achieve this, our approach utilizes both the post and reply texts as input for fine-tuning. The objective is to increase the vector similarity between a post and its genuine reply, while simultaneously decreasing the similarity for non-corresponding replies, which effectively serve as pseudo "Impression Zombies" replies.

\subsubsection{Classification with MLP Classifier}
\label{subsubsec:MLP}
To classify Impression Zombies, we construct a multifaceted input vector designed to capture the complex relationship between a post and its reply. This vector is formed using the embeddings generated by the fine-tuned model. Specifically, we concatenate four components: the two text embeddings from the model, their element-wise difference, and their element-wise product. This structure allows the final classifier to consider not only the texts' semantic similarity but also their shared features and differences.

\subsection{Experimental Setting}
 To verify the classification capabilities of our proposed model, we compared its classification performance with several existing models on the \textit{Classification Dataset}. For this experiment, the dataset was randomly split into a training set (80\%) and a test set (20\%). We employ the ``sbintuitions/sarashina-embedding-v1-1b''\footnote{https://huggingface.co/sbintuitions/sarashina-embedding-v1-1b} pre-trained language model as our embedding model. The hyperparameters for fine-tuning of SentenceTransformers were configured as follows: a loss function of MultipleNegativesRankingLoss, a batch size of 16, 8 training epochs, and a learning rate of 1e-5. We used three evaluation metrics: Precision, Recall, and Accuracy.

\subsection{Comparative Methods}
We evaluated the performance of our proposed model against several baselines: Logistic Regression, BERT, and GPT-4.1. We also examined the effect of fine-tuning on the model's accuracy.

\subsubsection{Logistic Regression}
The classification was performed on the parent post and reply pairs within the \textit{Classification dataset}. The input features for the model were created by first concatenating the text of a parent post with its corresponding reply, and subsequently transforming the combined text into a numerical vector using the Term Frequency-Inverse Document Frequency (TF-IDF) method.

\subsubsection{BERT}
The classification was performed on the parent post and reply pairs within the \textit{Classification dataset}. A pre-trained Japanese BERT model (tohoku-nlp/bert-base-japanese-whole-word-masking\footnote{https://huggingface.co/tohoku-nlp/bert-base-japanese-whole-word-masking}) was fine-tuned on the training data. The hyperparameters were configured as follows: a batch size of 8, 3 training epochs, and a learning rate of 2e-5. 

\subsubsection{GPT-4.1}
Using text pairs of parent posts and replies as input from the \textit{Classification Dataset}, we performed a classification. We provided the GPT-4.1 model with prompts consisting of parent post and reply pairs, and tasked the model with classifying the reply based on whether it had a semantic connection to its corresponding parent post. The experiment also compared the accuracy of zero-shot and few-shot prompting. In the few-shot setting, we provided a total of 10 examples: five replies from general users and five from Impression Zombies.

\subsection{Results}
\label{subsec:result}
\begin{table*}[t]
    \centering
    \caption{Classification accuracy results for each model. The labels ``General'' and ``Zombies'' correspond to ``general users'' and ``Impression Zombies,'' respectively.}
    \begin{tabular}{c|c|cccc}
       model & label & Precision & Recall & Accuracy\\
       \hline
       \multirow{2}{*}{Logistic Regression}  & General & 0.77 & 0.69 &  \multirow{2}{*}{0.71}\\
        & Zombies & 0.65 & 0.74 &  \\
        \hline
        \multirow{2}{*}{BERT} & General & \textbf{0.91} & 0.88 & \multirow{2}{*}{0.89}\\
        & Zombies & 0.85 & \textbf{0.90} & \\
        \hline
        \multirow{2}{*}{GPT-4.1 (Zero-shot)} & General & 0.71 & 0.73 &  \multirow{2}{*}{0.67}\\
        & Zombies & 0.63 & 0.60 & \\
        \hline
        \multirow{2}{*}{GPT-4.1 (Few-shot)} & General & \textbf{0.91} & 0.71 & \multirow{2}{*}{0.80}\\
        & Zombies & 0.71 & \textbf{0.90} & \\
        \hline
        \multirow{2}{*}{\textbf{Proposed Model (w/o fine-tuning)}} & General & 0.89 & \textbf{0.95} & \multirow{2}{*}{0.91}\\
        & Zombies & \textbf{0.93} & 0.85 & \\
        \hline
        \multirow{2}{*}{\textbf{Proposed Model (w/ fine-tuning)}} & General & \textbf{0.91} & \textbf{0.95} & \multirow{2}{*}{\textbf{0.92}}\\
        & Zombies & \textbf{0.93} & 0.88 & \\
        \hline
        
    \end{tabular}
    \label{tab:classify}
\end{table*}

Table \ref{tab:classify} shows the results of classifying replies from general users and Impression Zombies. The results indicate that our proposed model achieved the highest accuracy compared to the other three models. Furthermore, fine-tuning the embedding model improved the Precision for the ``general'' class and the Recall for the ``Zombies'' class, suggesting enhanced sensitivity in detecting Impression Zombies replies. However, the Precision for the ``Zombies'' class and the Recall for the ``general'' class did not increase, and the overall enhancement in accuracy was negligible. 

To further investigate model behavior, we conducted an error analysis. We found that many of the false positive pairs consisted of short yet coherent sentences. A characteristic of Impression Zombies replies is the frequent use of formulaic, machine-like text, and it is possible that these short replies were misclassified due to this perceived similarity. In contrast, false negative cases frequently involved replies that were exact duplicates of the parent post. Although this perfect textual match is unnatural human behavior, the model's reliance on semantic similarity caused it to interpret the resulting high textual match as high contextual coherence, thus leading to a false negative classification. These findings indicate that while contextual relationship is a valuable feature, the reliance of  solely on semantic coherence can introduce a bias against short yet valid replies. Furthermore, this approach fails to correctly identify cases such as duplications or repetitions, where textual similarity is high but the exchange is not conversationally coherent.

\section{Limitations}
This study has three main limitations. 
The first is the dataset size. This study utilized 9,909 replies and data from 5,497 users. Although it would have been desirable to acquire a larger dataset for more robust statistical analysis and model training.
Second, there is a temporal discrepancy between the fine-tuning data and the training/test data used for Impression Zombies classification. The data used to fine-tune the language model was collected more than four years before the data used for the actual classification. Consequently, the fine-tuning may not have been optimal for the time period in which the \textit{Parent-Reply Pair dataset} was collected, and as a result, may not have fully captured contemporary posting patterns.
Third, our classification model relies solely on text. Since we constructed a text-dependent model, replies made to images were not sufficiently considered. To improve classification accuracy further, it requires the integration of multimodal information, such as images, into the training process.

\section{Conclusion}
In this paper, we conducted a feature analysis of Impression Zombies using account data and performed classification of Impression Zombies using parent posts and their replies.
The feature analysis revealed the characteristics of Impression Zombies compared to general users. These included the possibility that they use the phrase ``follow back'' to increase their follower count and that their peak posting frequency occurs several hours after that of general users.
Furthermore, we confirmed that in our proposed model, fine-tuning the SentenceTransformers model enhances its ability to identify Impression Zombies replies, achieving a classification accuracy of 92\%. The findings obtained in this study are expected to contribute to the classification of Impression Zombies and to the improved integrity of information circulation on social networking services.

\section*{Acknowledgment}

This research was supported by a Grant-in-Aid for Scientific Research (Grant Number JP23K16889) from the Japan Society for the Promotion of Science (JSPS) and Research Institute of Science and Technology for Society (Grant Number JPMJRS23L4).

\bibliographystyle{plain}
\bibliography{myrefs}

\end{document}